# A Non-Habituating Configurable Audio Visual Animal Deterrent System to Mitigate Roadkill

Vedant Srinivas

*Eastlake High School*
*Sammamish, Washington, United States of America*
`vedant.m.srinivas@gmail.com`

*Abstract*— Over 30,000 miles of road are added yearly to the already enormous road system that exists in the United States. As roads segment habitats, animals have no option but to walk across them for food, water and companionship. In this process of accessing their basic needs, they end up becoming roadkill. Wherever wildlife habitat and roadways overlap, animal vehicle conflicts seem impossible to control. These animal deaths have a direct impact on the biodiversity and dynamics of an ecosystem and roadkill poses a threat to many species that are fighting extinction. Vehicles colliding with animals results in human fatalities, life changing injuries and extensive property damage. Current methods of handling roadkill are primarily passive and do not utilize the animal's natural instincts. This paper presents an alternative approach by actively involving the animal, warning them of an oncoming vehicle and triggering their survival instincts. Making the animal an integral part of the solution, augmenting their sensory perception with science and technology and utilizing their heightened reflexes and survival instincts provides a better chance at mitigating the problem of roadkill. The results show that this solution is able to provide animals a warning of an oncoming vehicle, consistently, reliably and in a wide range of testing conditions.

## I. INTRODUCTION

Most ungulates are crepuscular animals, i.e., are most active at dawn and dusk. Owing to this behavior, their visual acuity is highly tuned for low light conditions. When they are near roads and a rapidly-moving vehicle with bright lights approaches them with little warning, they are blinded and either freeze all movement or behave irrationally. Humans have comparatively slow reflexes in relation to animals, and braking at high speeds to avoid animal encounters leads to around 200 human fatalities and tens of thousands of injuries to drivers yearly. The total cost of damage to vehicles in these collisions reaches roughly $3.6 billion annually (Gaskill, 2013).

There are a total of 4.18 million miles of road in the United States (Russell, 2020). Additionally, there were 285 million vehicles registered in the USA in 2019 (US VIO Vehicle Registration Statistics, 2021). Due to this growth of roadways with primarily passive controls for animal protection it has reached a point where more than a million animal lives are lost on the roads of the United States every day (Roadkill Statistics, 2005).

Prevalent solutions to roadkill have relied on signage and good intentions of the driver. By making animals an integral part of the solution, augmenting their sensory perception with science and technology, and utilizing their heightened reflexes and natural survival instincts, this solution stands a better chance at mitigating the problem of roadkill.

Animals have always relied on sounds and road vibrations along with their sharp reflexes to avoid danger. The four noise generating components of a vehicle include; the engine, the exhaust, the aerodynamics, and the tires moving on the roads (Bernhard & Sandberg, 2005). With the rise of aerodynamically efficient electric drivetrains, low noise and vibration free roads the era of mobility is much less noisy than gasoline-powered vehicles. Distracted driving owing to a feature rich dashboard in addition to mobile device usage reduces driver attention and doubles human reaction times (Yager et al., 2012). With rapid urbanization and growth in population, cities have encroached on rural-urban interfaces, making the nation's roads deadly for wild animals.

Road mortality poses a threat to 21 endangered or threatened species including red wolves, certain species of deer, tortoises and even crocodiles in the United States alone ("Wildlife-Vehicle Collision Reduction Study: Report To Congress," 2008). Overall, roadkill has a drastic impact on the environment and poses many ecological risks, including threatening the biodiversity of ecosystems and extinction (Coffin, 2007).

## II. Existing Solutions

Passive road signage warning of animal presence in the area is one of the most common, and at the same time most ineffective mitigation measures (Rytwinski et al., 2016). There are other active deterrent solutions that have been tried and tested, but fail to show promise due to high costs and/or ineffectiveness and failure to address habituation concerns (Coulson & Bender, 2020).

Fencing is a roadkill reduction mechanism but requires a lot of planning and takes refined installation to get right. It is either too expensive when executed for millions of miles of highways or is just not feasible in high density areas. When installed, care must be taken to avoid creating a permanent divider that keeps animals from accessing habitat on the other side of the fence ("Wildlife-Vehicle Collision Reduction Study: Report To Congress.," 2008). If easy opportunities for accessing habitats across roads (through gaps in fencing or bridges and tunnels) are not made available, animals will try to break through the fencing and this in turn causes more financial drain through operations and maintenance work. Bridges and underpasses are possible solutions to provide this safe crossing but they are very expensive and there is no way to force the animals along these paths without accompanying fencing.

An active driver warning system warns drivers of animal presence in the area. The driver becomes aware of the animal's potential presence, but combining the time it takes for the average driver to react to the animal on the road, apply the brakes, and come to a complete stop, would mean a car traveling at 60 mph would travel over 270 feet in ideal driving conditions. Such sudden breaking also leads to loss of control and accidents for the driver, other vehicles nearby, and potential property damage.

## III. Proposed Solution

The most prevalent solutions to roadkill are primarily passive. In order to make the animal an integral part of the solution, this device senses an oncoming vehicle and alerts an animal through audio and visual cues. This enables utilizing the animal's relatively quick reaction times and safety instinct to help prevent the loss of lives and property damage.

In order to accomplish the goal of getting the animal out of harm's way, two things are needed. First is a device that can sense and provide the warning to the animal with enough time to act on it and the second is to deliver the trigger in the most effective way to induce the right response. There is existing research in identifying the triggers that are most effective in invoking the animal's escape response. So, if successful in triggering accurately with enough time to act, then the solution is successful.

The proposed solution is a device that detects vehicles by their headlights, utilizing ambient light sensors along with self learning algorithms to detect the luminosity in either direction of the road. If this luminosity is increasing at a rapid pace then a warning light starts to blink and the system starts playing a random deterring sound. This audio and visual signal is to alert any animals in the near vicinity, allowing them to get out of the oncoming automobile's path early, while the blinking lights warn the driver of an animal prone stretch of road.

## IV. Method

*Components*

This device is made up of several components that have been integrated into a custom designed printed circuit board. These components include, an Arduino MkrZero Board, two I2C ambient light sensor circuits, a multiplexer, a solar charge controller, a Li-Po battery, a solar panel, and an I2S circuit containing an 8 ohm speaker. The device is driven by an intelligent, self learning and calibrating algorithm that can minimize false positives, have high recall, precision, and low trigger latency under a wide range of lighting conditions. The device cycles through multiple deterrent sounds to avoid habituation of the animals to any one trigger.

*Noise Filtering*

Ambient Light Sensors (ALS) provide measurements of ambient light intensity which match the human eye's response to light under a variety of lighting conditions. Each sensor has a specific operating range of performance, from very low light up to bright sunlight. The ALS that was used for this research has a detection range between 188 uLux and 88000 Lux. This wide dynamic range made this ALS ideal for the operating scenario between midnight dark and oncoming headlights. The system uses the ratio of the sensor readings to the baseline light reading to identify an approaching vehicle. The raw signal generated by the ALS sensor is noisy and in its raw form generates a high false positive rate.

A simple moving average algorithm is good at removing noise in data, but when run through the ALS signal it was unable to effectively filter out the false positives. In order to eliminate noise effectively, an inertial moving average algorithm was developed.

$$IMA_i = IMA_{i-1} \cdot \alpha + (1 - \alpha) \cdot A_i$$

$A_i$ is the curent sensor value. Changing $\alpha$ changes the inertia.

This algorithm keeps track of the moving value of the sensor reading. With every new data point, the algorithm weighs the moving value by the inertia parameter and the complement of the weight for the latest reading. This filters out high noise levels and provides a configurable model for noise vs signal reactiveness.

*Calibration and Change Detection*

In order to detect when the light changes at a quick rate the algorithm maintains two Inertial moving average values. The first value has very high inertia and moves slowly - this is the baseline light level. The second inertial moving average has low inertia and follows the raw signal closely while filtering out the noise - this is called the instantaneous light level. These signal plots for a period of time are shown in Figure 1 with a simple moving average and the raw signal as comparison data points. The blue line represents the raw signal and yellow the baseline IMA.

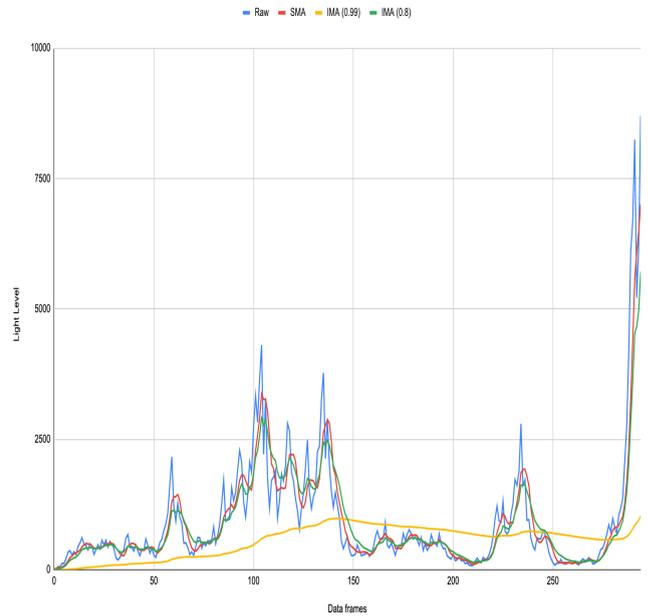

Fig 1: Relationship between time and various filtering algorithms.

When the ratio of the instantaneous to baseline exceeds a specific threshold the system triggers a positive detection. The Inertial Moving Average algorithm also adapts to the light level of the environment and prevents the need to globally normalize every sensor's signal range as well as calibrate to deployment site conditions.

*Habituation*

Habituation is a common problem faced by technology when addressing wildlife related issues. Animals start to get used to deterrents when there is no positive or negative impact afterwards. (Coulson & Bender, 2020). Evidence exists of fallow deer getting habituated to similar tonal sounds after 5 days of testing (Ujvári et al., 2004). A similar experiment run with roe deer demonstrated that natural sounds like wolf howls improved reaction times and proved much more useful, while also remaining just as effective throughout the 5 years of the study (Babińska-Werka et al., 2015).

This device was built to take in an SD card with audio files of the sounds that could be used as triggers. The trigger sounds are selectively curated and tuned for common animals in a specific zone.

The device, when triggered, plays these sounds randomly to avoid habituation to any one sound.

From a visual acuity perspective, deer, one of the leading victims of roadkill, are more sensitive to certain colors. White-tailed and fallow deer had a high range of sensitivity with light wavelengths from 450 to 542 nanometers, which represent shades of blue and green (Jacobs et al., 1994). As deer represent a large range of ungulates, which are highly susceptible to roadkill, the choice of the visual signal was kept to be configurable and defaulted to the green spectrum (for the most common animal species in the study area).

## Testing & Results

The device was set up on a long stretch of a four lane state highway. The ambient lighting at the time of study was in line with night time driving conditions. The speed limit on this road was 45 mph. The vehicle type was noted for post analysis. Other variability was introduced by testing under street lights as well as completely dark conditions.

TABLE I
END TO END PERFORMANCE

| Metric | Performance |
| --- | --- |
| Recall | 96.5% |
| Precision | 100% |
| False Positive Rate | 0% |

The device was tested on 200 random instances of oncoming vehicles. In 193 of those instances the vehicles were successfully identified and the audio visual trigger was engaged, yielding a recall of 96.5% at a precision of 100% and a false positive rate of 0%. The device was able to provide an average trigger of 3.7 seconds and a trigger distance of 244 ft for the animal to react, and for bigger vehicles this number went up to over 500 ft, yielding 10 seconds for the animal to get out of harm's way.

## Discussion

A histogram of detections is presented in Figure 2. Any vehicle that was detected within 60 ft of the device was considered to be a failure as it would not provide enough time for the animal to react. In real world settings this trigger is still useful for animals downstream from the device location.

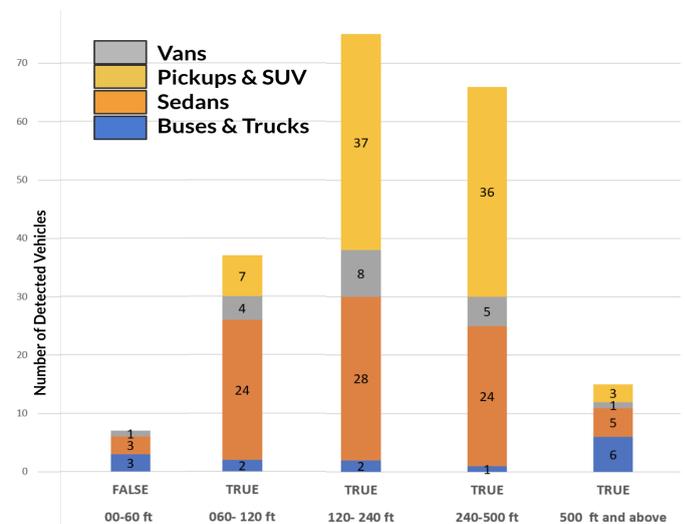

Fig. 2 Histogram of vehicle detections by type

Four of the failures were with old vehicles having yellow, incandescent headlights. 14% of the vehicles fell into this category, and the average distance triggers were 157 ft., a 39% reduction compared to the overall average. The other three false negatives were with semi trucks that were in the farthest lane from the device and were following a caravan of vehicles, which had continuously triggered the device for a long period of time. When there is continuous high intensity light for extended periods of time, the system stops triggering to avoid noise and light pollution. Even though this is by design for the purpose of this research these data points were considered failures.

Since most vehicles travel with low beams, placing them at about 3-4 Ft from the ground yielded the greatest trigger distance. Buses and Trucks have a higher center for their light cones and were able to trigger over long distances even when the device was placed at heights over 4 ft.

On highways with many lanes (greater than 4), or those which have tall dividers along the middle, light from one side of the freeway might not reach the other side. The light cone of vehicles being narrow to avoid distraction for traffic coming from the other direction coupled with road dividers that are meant to ensure crossing traffic safety are challenges for the device. The recall and precision numbers remained unchanged for up to 4 lanes of road. In scenarios where there are more than 4 lanes, it is better to install the devices in the middle of the freeway having equal visibility to both sides.

The device is able to achieve the goals of recall, precision and false positive rate necessary to be effective and reliable. As next steps the device will need to be installed in stretches of roads having prevalent roadkill alongside uninstrumented stretches of roads and a comparison test will be performed to document the complete effectiveness of the solution in mitigating roadkill.